\documentclass[12pt]{article}
\usepackage{amssymb,amsmath,amscd,latexsym,graphicx,epsfig}
\begin{document}

\title{The Strong Free Will Theorem}

\setcounter{footnote}{0}
\renewcommand{\thefootnote}{\fnsymbol{$\ast$}}

\author{John Conway$^\ast$ and Simon Kochen\footnote{jhorcon@yahoo.com;
kochen@math.princeton.edu}\\
Princeton University\\
Department of Mathematics \\
Princeton, NJ 08544--1000} 
%email:  {\it jhorcon\@yahoo.com; kochen\@math.princeton.edu}}

\date{June 12, 2008}

%begin{document}

%\abstract
%\endabstract

\maketitle

\parskip =.3cm
\parindent=.5cm
\baselineskip=12pt

\section{Introduction}
%1. Introduction

\indent \hskip .4cm The two theories that revolutionized physics 
in the 20th century, relativity and quantum mechanics, are full of 
predictions that defy common sense. Recently, we used three such 
paradoxical ideas to prove ``The Free Will Theorem" (strengthened here), 
which is the culmination of a series of theorems about quantum mechanics 
that began in the 1960's.  It asserts, roughly, that if indeed we humans 
have free will, then elementary particles already have their own small 
share of this valuable commodity.  More precisely, if the experimenter 
can freely choose the directions in which to orient his apparatus in a 
certain measurement, then the particle's response (to be pedantic -- the 
universe's response near the particle) is not determined by the entire 
previous history of the universe.

Our argument combines the well-known consequence of relativity theory,
that the time order of space-like separated events is not absolute, with
the EPR paradox discovered by Einstein, Podolsky and Rosen in 1935, and
the Kochen-Specker Paradox of 1967 (See [2].)  We follow Bohm in using a
spin version of EPR and Peres in using his set of 33 directions, rather
than the original configuration used by Kochen and Specker. More
contentiously, the argument also involves the notion of free will, but
we postpone further discussion of this to the last section of the paper.

Note that our proof does not mention ``probabilities" or the
``states" that determine them, which is fortunate because these
theoretical notions have led to much confusion. For instance, it is
often said that the probabilities of events at one location can be
instantaneously changed by happenings at another space-like separated
location, but whether that is true or even meaningful is irrelevant 
to our proof, which never refers to the notion of probability.

For readers of the original version [1] of our theorem, we note that
we have strengthened it by replacing the axiom FIN together with the
assumption of the experimenters' free choice and temporal causality by a
single weaker axiom MIN. The earlier axiom FIN of [1], that there
is a finite upper bound to the speed with which information can be
transmitted, has been objected to by several authors. Bassi and Ghirardi
asked in [3]: what precisely is ``information," and do the ``hits" and
``flashes" of GRW theories (discussed in the Appendix) count as
information? Why cannot hits be transmitted instantaneously, but not
count as signals? These objections miss the point. The only information
to which we applied FIN is the choice made by the experimenter and the
response of the particle, as signaled by the orientation of the
apparatus and the spot on the screen. The speed of transmission of any
other information is irrelevant to our argument. The replacement of FIN
by MIN has made this fact explicit. The theorem has been further
strengthened by allowing the particles' responses to depend on past
half-spaces rather than just the past light cones of [1].

\section{The Axioms}
%2. The Axioms.

\indent \hskip .4cm We now present and discuss the three axioms on which 
the theorem rests.

(i) The SPIN Axiom and the Kochen-Specker Paradox.

Richard Feynman once said that ``If someone tells you they understand
quantum mechanics, then all you've learned is that you've met a liar."
Our first axiom initially seems easy to understand, but beware --
Feynman's remark applies! The axiom involves the operation called
``measuring the squared spin of a spin $1$ particle," which always produces
the result $0$ or $1$.

SPIN Axiom: {\em Measurements of the squared (components of) spin of
a spin 1 particle in three orthogonal directions always give the answers
$1,0,1$ in some order.}

Quantum mechanics predicts this axiom since for for a
spin 1 particle the squared spin operators  $s_x^2, s_y^2, s_z^2$
commute and have sum 2.

This ``$101$ property" is paradoxical because it already implies that
the quantity that is supposedly being measured cannot in fact exist
before its ``measurement."
For otherwise there would be a function defined on the sphere of
possible directions taking each orthogonal triple to $1,0,1$ in some
order. It follows from this that it takes the same value on pairs of
opposite directions, and never takes two orthogonal directions to $0$.

We call a function defined on a set of directions that has all three
of these properties a ``$101$ function" for that set.  But unfortunately
we have:

The Kochen-Specker Paradox: {\em there does not exist a $101$
function for the $33$ pairs of directions of Figure 1 (the Peres
configuration).}

\vskip .6cm

\begin{figure}[!ht]
\centerline{
\epsfxsize=0.4in
\epsfysize=0.2in}
\includegraphics{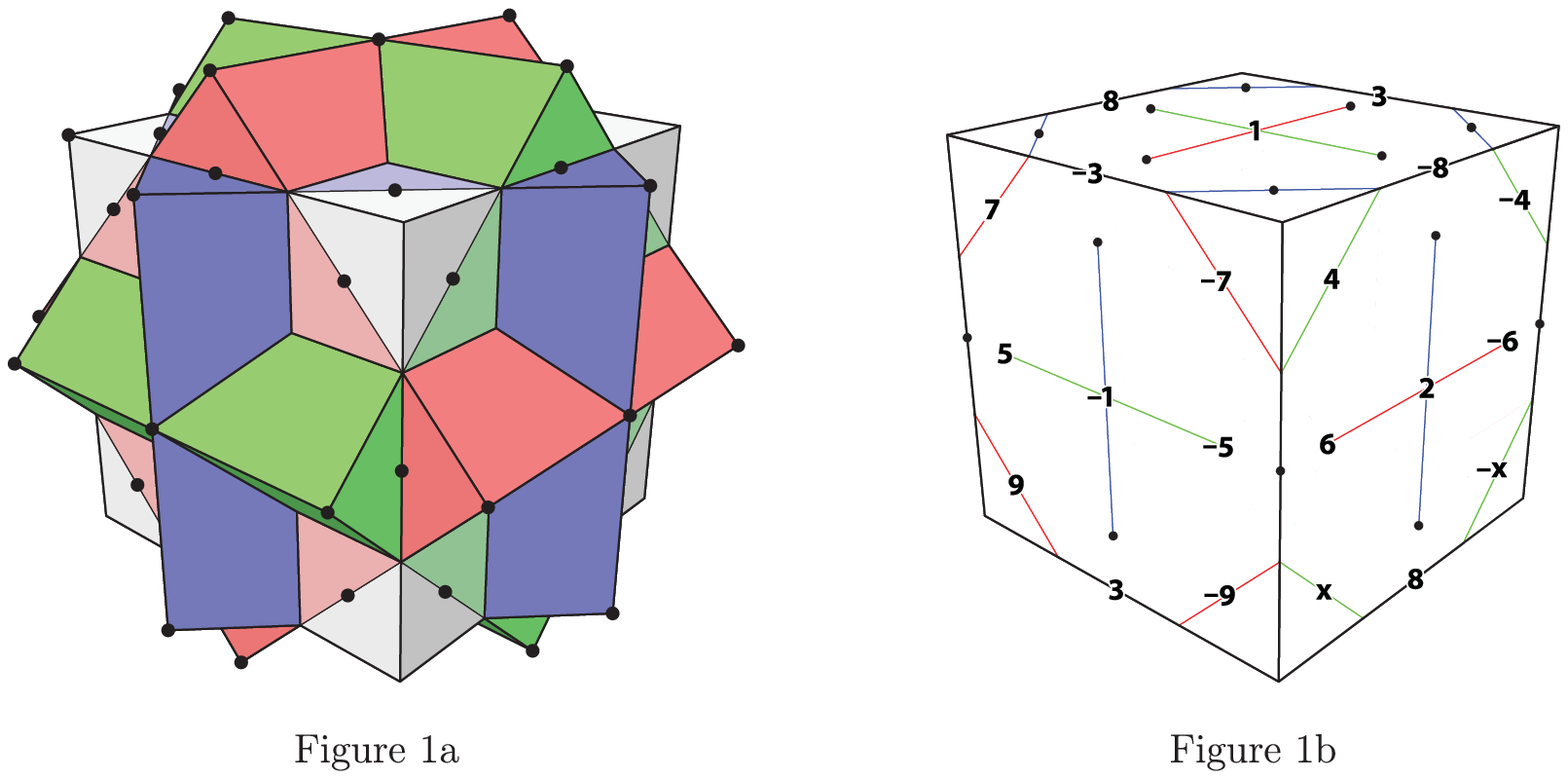}
\caption{\footnotesize{The three colored cubes in Figure 1a are obtained by 
rotating the white cube through $45^\circ$ about its 
coordinate axes. The 33 directions are the symmetry axes of the 
colored cubes, and pass through the spots in Figure 1a. 
Figure 1b shows where these directions meet the white cube.}}
\end{figure}

\pagebreak

\noindent {\it Proof}.  We shall call a node even or odd according 
as the putative $101$ function is supposed to take the value $0$ or $1$ at it, 
and we progressively assign even or odd numbers to the nodes in Figure 1b 
as we establish the contradiction.

We shall use some easily justified orthogonalities -- for instance the 
coordinate triple rotates to the triple $(2,3,-3)$ that starts our proof, 
which in turn rotates (about $-1$) to the triples $(8,-7,9)$ and 
$(-8,7,-9)$ that finish it.

Without loss of generality nodes $1$ and $-1$ are odd and node $2$ even,
forcing $3$ and $-3$ to be odd.  Now nodes $4$ and -x form
a triple with $3$, so one of them (w.l.o.g. $4$) is even.
In view of the reflection that interchanges
$-4$ and x while fixing $4$ and -x, we can w.l.o.g. suppose
that $-4$ is also even. 

There is a $90^\circ$ rotation about $1$ that moves $7$, $5$, $9$ to
$4$, $6$, x, showing that $5$ is orthogonal to $4$, while
$1$, $5$, $6$ is a triple, and also that $6$ is orthogonal to both
$7$ and $9$.  Thus $5$ is odd, $6$ even, and $7$, $9$ odd.
A similar argument applies to nodes $-5$, $-6$, $-7$, $-9$.  

Finally, $8$ forms a triple with $-7$ and $9$, as does $-8$ with $7$ and
$-9$.  So both these nodes must be even, and since they are orthogonal,
this is a contradiction that completes the proof. \newline
\indent \hskip 6.2in $\square$

Despite the Kochen-Specker paradox, no physicist would question the 
truth of our SPIN axiom, since it follows from quantum mechanics, which 
is one of the most strongly substantiated scientific theories of all 
time. However, it is important to realize that we do not in fact suppose 
all of quantum mechanics, but only two of its testable consequences, 
namely this axiom SPIN and the axiom TWIN of the next section.

It is true that these two axioms deal only with idealized forms of
experimentally verifiable predictions, since they refer to exact
orthogonal and parallel directions in space. However, as we have shown
in [1], the theorem is robust in that approximate forms of these axioms
still lead to a similar conclusion. At the same time, this shows that
any more accurate modifications of special relativity (such as general
relativity) and of quantum theory will not affect the conclusions of the
theorem.

(ii) The TWIN Axiom and the EPR Paradox.

One of the most curious facts about quantum mechanics was pointed
out by Einstein, Podolsky and Rosen in 1935.  This says that even
though the results of certain remotely separated observations cannot
be individually predicted ahead of time, they can be correlated.

In particular, it is possible to produce a pair of ``twinned" spin 1 particles
(by putting them into the ``singleton state" of total spin zero)
that will give the same answers to the above squared spin measurements
in parallel directions. Our ``TWIN" axiom is part of this assertion.

The TWIN Axiom: {\em  For twinned spin 1 particles, suppose experimenter
$A$ performs a triple experiment of measuring the squared spin component
of particle $a$ in three orthogonal directions $x,y,z$, while
experimenter $B$  measures the twinned particle $b$ in one direction,
$w$. Then if $w$ happens to be in the same direction as one of $x,y,z$,
experimenter $B$'s measurement will necessarily yield the same answer as 
the corresponding measurement by $A$.}

In fact we will restrict $w$ to be one of the 33 directions in the
Peres configuration of the previous section, and $x,y,z$ to be one of
40 particular orthogonal triples, namely the 16 such triples of that
configuration and the 24 further triples obtained by completing its
remaining orthogonal pairs.

(iii) The MIN Axiom, Relativity, and Free Will.

One of the paradoxes introduced by relativity was the fact that
temporal order depends on the choice of inertial frame. If two events
are space-like separated, then they will appear in one time order with
respect to some inertial frames, but in the reverse order with respect
to others. The two events we use will be the above twinned spin
measurements.

It is usual tacitly to assume the temporal causality principle that
the future cannot alter the past. Its relativistic form is that an event
cannot be influenced by what happens later in any given inertial frame.
Another customarily tacit assumption is that experimenters are free to
choose between possible experiments.  To be precise, we mean that the
choice an experimenter makes is not a function of the past. We
explicitly use only some very special cases of these assumptions in
justifying our final axiom.

The MIN Axiom: {\em Assume that the experiments performed by $A$ and $B$ 
are space-like separated. Then experimenter $B$ can freely choose any
one of the 33 particular directions $w$, and $a$'s response is
independent of this choice. Similarly and independently, $A$ can freely
choose any one of the 40 triples $x,y,z$, and $b$'s response is
independent of that choice.}

It is the experimenters' free will that allows the free and independent
choices of $x,y,z$ and $w$.  But in one inertial frame -- call it the
``$A$-first" frame -- $B$'s experiment will only happen some time later
than $A$'s, and so $a$'s response cannot, by temporal causality, be
affected by $B$'s later choice of $w$.  In a $B$-first frame, the 
situation is reversed, justifying the final part of MIN. (We shall 
discuss the meaning of the term ``independent" more fully in the Appendix.)

\section{The (Strong) Free Will Theorem}
%3. The (Strong) Free Will Theorem

\indent \hskip .4cm Our theorem is a strengthened form of the original 
version of [1].  Before stating it, we make our terms more precise. 
We use the words ``properties," ``events" and ``information"  almost
interchangeably: whether an event has happened is a property, and
whether a property obtains can be coded by an information-bit. The exact
general meaning of these terms, which may vary with some theory that may
be considered, is not important, since we only use them in the specific
context of our three axioms.

To say that $A$'s choice of $x,y,z$ is free means more precisely that
it is not determined by (i.e., is not a function of) what has happened
at earlier times (in any inertial frame).  Our theorem is the surprising
consequence that particle $a$'s response must be free in exactly the
same sense, that it is not a function of what has happened earlier (with
respect to any inertial frame).

\noindent{\bf The Free Will Theorem}. 
{\em The axioms SPIN, TWIN and MIN imply that
the response of a spin 1 particle to a triple experiment is free --- that
is to say, is not a function of properties of that part of the universe
that is earlier than this response with respect to any given inertial
frame.}

\noindent {\it Proof}.  We suppose to the contrary -- this is the ``functional 
hypothesis" of [1] -- that particle $a$'s response $(i,j,k)$ to the 
triple experiment with directions $x,y,z$ is given by a function of 
properties $\alpha ,\ldots$ that are earlier than this response
with respect to some inertial frame $F$. We write this as
$$\theta_a^F(\alpha) = \text{one of } (0,1,1), (1,0,1), (1,1,0)$$
(in which only a typical one of the properties $\alpha$ is indicated).

Similarly we suppose that $b$'s response $0$ or $1$ for the direction 
$w$ is given by a function
$$\theta_b^G(\beta) = \text{ one of } 0 \text{ or } 1$$
of properties $\beta,...$ that are earlier with respect to a possibly
different inertial frame $G$.

(i) If either one of these functions, say $\theta_a^F$, is influenced by
some information that is free in the above sense (i.e., not a function
of $A$'s choice of directions and events F-earlier than that choice),
then there must be an an earliest (``infimum") F-time $t_0$ after
which all such information is available to $a$. Since the non-free
information is also available at $t_0$, all these information bits, free
and non-free, must have a value $0$ or $1$ to enter as arguments in the
function $\theta_a^F$. So we regard $a$'s response as having started at $t_0$.

If indeed, there is {\em any} free bit that influences $a$, the
universe has by definition taken a free decision near $a$ by time $t_0$,
and we remove the pedantry by ascribing this decision to particle $a$.
(This is discussed more fully in \S $4$.)

(ii) From now on we can suppose that no such new information bits
influence the particles' responses, and therefore that $\alpha$ and
$\beta$ are functions of the respective experimenters' choices and
of events earlier than those choices.

Now an $\alpha$ can be expected to vary with $x,y,z$ and may or may not
vary with $w$. However, whether the function varies with them or not, we
can introduce all of $x,y,z,w$ as new arguments and rewrite $\theta_a^F$
as a new function (which for convenience we give the same name)
$$\theta_a^F(x,y,z,w;\alpha')   \eqno{(\star)} $$
of $x,y,z,w$ and properties $\alpha'$ independent of $x,y,z,w$.

To see this, replace any $\alpha$ that does depend on $x,y,z,w$ by the
constant values $\alpha_1, \ldots , \alpha_{1320}$ it takes for 
the $40 \times 33 = 1320$ particular quadruples $x,y,z,w$ we shall use. 
Alternatively, if each $\alpha$ is some function $\alpha(x,y,z,w)$ 
of $x,y,z,w$, we may substitute these functions in $(\star)$ to 
obtain information bits independent of $x,y,z,w$.

Similarly, we can rewrite $\theta_b^G$ as a function
$$\theta_b^G(x,y,z,w;\beta')$$
of $x,y,z,w$ and properties $\beta'$ independent of $x,y,z,w$.

Now there are values $\beta_0$ for $\beta'$, for which
$$\theta_b^G(x,y,z,w;\beta_0)$$
is defined for whatever choice of $w$ that $B$ will make, and therefore,
by MIN, for all the 33 possible choices he is free to make at that
moment (since $B$ can choose independently of $\beta$).

We now define
$$\theta_0^G(w) = \theta_b^G(x,y,z,w;\beta_0) \,,$$
noting that since by MIN the response of $b$ cannot vary with $x,y,z$,
$\theta_0^G$ is a function just of $w$.

Similarly there is a value $\alpha_0$ of $\alpha'$ for which the function
$$\theta_1^F(x,y,z) = \theta_a^F(x,y,z,w;\alpha_0)$$
is defined for all 40 triples $x,y,z$, and it is also independent of
$w$, which argument we have therefore omitted.

But now by TWIN we have the equation
$$\theta_1^F(x,y,z) = (\theta_0^G(x),\; \theta_0^G(y), \; \theta_0^G(z) ) \,.$$
However, since by SPIN the value of the left-hand side is one of
$(0,1,1)$, $(1,0,1)$, $(1,1,0)$, this shows that $\theta_0^G$ is a 101
function, which the Kochen-Specker paradox shows does not exist. This
completes the proof.  \hskip 4.512in $\square$

\section{Locating the Response}
%4. Locating the Response

\indent \hskip .4cm We now provide a fuller discussion of some delicate points.

(i) Since the observed spot on the screen is the result of a cascade of
slightly earlier events, it is hard to define just when ``the response"
really starts. We shall now explain why one can regard $a$'s response
(say) as having already started at any time after $A$'s choice when all
the free information bits that influence it have become available to $a$.

Let $N(a)$ and $N(b)$ be convex regions of space-time that are
just big enough to be ``neighborhoods of the respective experiments", by
which we mean that they contain the chosen settings of the apparatus and
the appropriate particle's responses. Our proof has shown that if
the backward half-space $t < t_F$ determined by a given F-time $t_F$ is
disjoint from $N(a)$, then the {\em available} information it contains
is not enough to determine $a$'s response. On the other hand, if each of
the two such half-spaces contains the respective neighborhood, then of
course they already contain the responses. By varying $F$ and $G$, this
suffices to locate the free decisions to the two neighborhoods, which
justifies our ascribing it to the particles themselves.

(ii) We remark that not all the information in the $G$-backward
half-space (say) need be available to $b$, because MIN prevents
particle $b$'s function $\theta_b^G$ from using experimenter $A$'s
choice of directions $x,y,z$. The underlying reason is of course, that
relativity allows us to view the situation from a $B$-first frame, in
which $A$'s choice is made only later than $b$'s response, so that $A$
is still free to choose an arbitrary one of the 40 triples. However,
this is our only use of relativistic invariance - the argument actually
allows any information that does not reveal $A$'s choice to be
transmitted superluminally, or even backwards in time.

(iii) Although we've precluded the possibility that $\theta_b^G$ can
vary with $A$'s choice of directions, it is conceivable that it might
nevertheless vary with $a$'s (future!) response. 
However, $\theta_b^G$ cannot be affected by $a$'s response to
an unknown triple chosen by $A$, since the same information is
conveyed by the responses $(0,1,1)$, to $(x,y,z)$, $(1,0,1)$ to
$(z,x,y)$, and $(1,1,0)$ to $(y,z,x)$.
For a similar 
reason $\theta_a^F$ cannot use $b$'s response, since $B$'s experiment 
might be to investigate some orthogonal triple $u,v,w$ and discard the 
responses corresponding to $u$ and $v$.

(iv) It might be objected that free will itself might in some sense be
frame-dependent.  However, the only instance used in our proof
is the choice of directions, which, since it becomes manifest in the
orientation of some macroscopic apparatus, must be the same as seen
from arbitrary frames.

(v)  Finally, we note that the new proof involves four
inertial frames -- $A$-first, $B$-first, $F$ and $G$. This number cannot
be reduced without weakening our theorem, since we want it to apply to
arbitrary frames $F$ and $G$, including for example those in which the
two experiments are nearly simultaneous. 

\section{Free Will versus Determinism}
%5. Free Will versus Determinism.

\indent \hskip .4cm We conclude with brief comments on some of the more 
philosophical consequences of the Free Will Theorem (abbreviated to FWT).

Some readers may object to our use of the term ``free will" to
describe the indeterminism of particle responses. Our provocative
ascription of free will to elementary particles is deliberate, since our
theorem asserts that if experimenters have a certain freedom, then
particles have exactly the same kind of freedom. Indeed, it is natural to
suppose that this latter freedom is the ultimate explanation of our own.

The humans who choose $x,y,z$ and $w$ may of course be replaced by a
computer program containing a pseudo-random number generator. If we
dismiss as ridiculous the idea that the particles might be privy to this
program, our proof would remain valid. However, as we remark in [1],
free will would still be needed to choose the random number generator, 
since a determined determinist could maintain that this choice was 
fixed from the dawn of time.

We have supposed that the experimenters' choices of directions
from the Peres configuration are totally free and independent.  However,
the freedom we have deduced for particles is more constrained, since it
is restricted by the TWIN axiom.  We introduced the term ``semi-free" in 
[1] to indicate that it is really the pair of particles that jointly 
makes a free decision.

Historically, this kind of correlation was a great surprise, which
many authors have tried to explain away by saying that one particle
influences the other.  However, as we argue in detail in [1], the
correlation is relativistically invariant, unlike any such explanation.
Our attitude is different:  following Newton's famous dictum ``Hypotheses
non fingo," we attempt no explanation, but accept the correlation as a
fact of life.

Some believe that the alternative to determinism is randomness,
and go on to say that ``allowing randomness into the world does not
really help in understanding free will.'' 
However, this objection does not apply
to the free responses of the particles that we have described. It may well
be true that classically stochastic processes such as tossing a
(true) coin do not help in explaining free will, but, as we show in
the Appendix and in \S10.1 of [1], adding randomness also does not
explain the quantum mechanical effects described in our theorem. It is
precisely the ``semi-free" nature of twinned particles, and more
generally of entanglement, that shows that something very different from
classical stochasticism is at play here. 

Although the FWT suggests to us that determinism is not a viable
option, it nevertheless enables us to agree with Einstein that 
``God does not play dice with the Universe.''
In the present state of knowledge, it is certainly beyond our 
capabilities to understand the connection between the free decisions
of particles and humans, but the free will of neither of these is 
accounted for by mere randomness.

The tension between human free will and physical determinism has a
long history. Long ago, Lucretius made his otherwise deterministic 
particles ``swerve" unpredictably to allow for free will. It was largely the 
great success of deterministic classical physics that led to the 
adoption of determinism by so many philosophers and scientists, 
particularly those in fields remote from current physics. (This remark 
also applies to ``compatibalism," a now unnecessary attempt
to allow for human free will in a deterministic world.)

Although, as we show in [1], determinism may formally be shown to
be consistent, there is no longer any evidence that supports it, in view
of the fact that classical physics has been superseded by quantum
mechanics, a non-deterministic theory. The import of the free will
theorem is that it is not only current quantum theory, but the world 
itself that is non-deterministic, so that no future theory can return us 
to a clockwork universe.

\section{Appendix. {\normalsize{Can there be a Mechanism for Wave
Function Collapse?}}}
%6. Appendix. Can there be a Mechanism for Wave Function Collapse?

\indent \hskip .4cm \footnotesize{Granted our three axioms, 
the FWT shows that Nature itself is non-deterministic.
It follows that there can be no correct relativistic deterministic 
theory of nature. In particular, no relativistic version of a hidden 
variable theory such as Bohm's well-known theory [4] can exist.

Moreover, the FWT has the stronger implication that there can be no
relativistic theory that provides a mechanism for reduction. There are
non-linear extensions of quantum mechanics, which we shall call
collectively GRW theories (after Ghirardi, Rimini, and Weber, see [5]) 
that attempt to give such a mechanism. The original theories
were not relativistic, but some newer versions make that claim. We shall
focus here on Tumulka's theory rGRWf (See [6]), but our argument
below applies, mutatis mutandis, to other relativistic GRW theories.
We disagree with Tumulka's claim in [7] that the FWT does not apply 
to rGRWf, for reasons we now examine.

(i) As it is presented in [6], rGRWf is not a deterministic theory.
It includes stochastic ``flashes" that determine the particles'
responses. However, in [1] we claim that adding randomness, or a
stochastic element, to a deterministic theory does not help:

``To see why, let the stochastic element in a putatively relativistic
GRW theory be a sequence of random numbers (not all of which need be
used by both particles). Although these might only be generated as
needed, it will plainly make no difference to let them be given in
advance. But then the behavior of the particles in such a theory would
in fact be a function of the information available to them (including
this stochastic element)."

Tumulka writes in [7] that this ``recipe" does not apply to rGRWf:

``Since the random element in rGRWf is the set of flashes, nature
should, according to this recipe, make at the initial time the decision
where-when flashes will occur, make this decision ``available" to every
space-time location, and have the flashes just carry out the
pre-determined plan. The problem is that the distribution of the flashes
depends on the external fields, and thus on the free decision of the
experimenters. In particular, the correlation between the flashes in
$A$ and those in $B$ depends on both external fields. Thus, to let the
randomness ``be given in advance" would make a big difference indeed, as
it would require nature to know in advance the decision of both
experimenters, and would thus require the theory either to give up
freedom or to allow influences to the past."

Thus, he denies that both our ``functional hypothesis," and therefore
also the FWT, apply to rGRWf . However, we can easily deal with the 
dependence of the distribution of flashes on the external fields $F_A$ 
and $F_B$, which arise from the two experimenters' choices  of 
directions $x,y,z$ and $w$. There are 
$40 \times 33 = 1320$ possible fields in 
question. For each such choice, we have a distribution $X(F_A,F_B)$ of 
flashes, i.e. we have different distributions $X_1,X_2,\ldots ,X_{1320}$. 
Let us be given ``in advance" all such random sequences, with their 
different weightings as determined by the different fields. Note that 
for this to be given, nature does not have to know in advance the actual 
free choices  $F_A$ (i.e. $x,y,z$) and $F_B$ (i.e. $w$) of the 
experimenters. Once the choices are made, nature need only refer to the 
relevant random sequence $X_k$ in order to emit the flashes in accord 
with rGRWf.

If we refer to the proof of the FWT, we can see that we are here simply
treating the distributions $X(F_A,F_B)$ $[= X(x,y,z,w)]$ in exactly
the same way we treated any other information-bit $\alpha$ that depended
on $x,y,z,w$. There we 
substituted all the values $\alpha_1, \ldots, \alpha_{1320}$
for $\alpha$ in the response function $\theta_a (x, y, z, w; \alpha)$.
Thus, the functional hypothesis does apply to rGRWf, as modified in this
way by the recipe.

Tumulka [7] grants that if that is the case, then rGRWf
acquires some nasty properties: In some frame $\Lambda$, ``[the flash]
$f_y^\Lambda$ will entail influences to the past." Actually, admitting
that the functional hypothesis applies to rGRWf has more dire
consequences -- it leads to a contradiction. For if, as we just showed,
the functional hypothesis applies to the flashes, and the first flashes
determine the particles' responses, then it also applies to these
responses which, by the FWT, leads to a contradiction.

(ii) Another possible objection is that in our statement of the MIN
axiom, the assertion that $a$'s response is independent of $B$'s choice
was insufficiently precise. Our view is that the statement must be true
whatever precise definition is given to the term ``independent," because
in no inertial frame can the past appearance of a macroscopic spot on a
screen depend on a future free decision.

It is possible to give a more precise form of MIN by replacing the
phrase ``particle $b$'s response is independent of $A$'s choice" by 
``if $a$'s response is determined by $B$'s choice, then its value does not
vary with that choice". However, we actually need precision only in the
presence of the functional hypothesis, when it takes the mathematical
form that $a$'s putative response function $\theta_a^F$ cannot in fact
vary with $B$'s choice. To accept relativity but deny MIN is therefore
to suppose that an experimenter can freely make a choice that will alter
the past, by changing the location on a screen of a spot that has
already been observed.

Tumulka claims in [7] that since in the twinning experiment the
question of which one of the first flashes at $A$ and $B$ is earlier is
frame dependent, it follows that the determination of which flash
influences the other is also frame dependent. However, MIN does not deal
with flashes or other occult events, but only with the particles'
responses as indicated by macroscopic spots on a screen, and these are
surely not frame dependent.

In any case, we may avoid any such questions about the term
``independent" by modifying MIN to prove a weaker version of the FWT,
which nevertheless still yields a contradiction for relativistic GRW
theories, as follows.

MIN$^\prime$:  In an $A$-first frame, $B$ can freely choose any one of the 33
directions $w$, and $a$'s prior response is independent of $B$'s choice.
Similarly, in a $B$-first frame, $A$ can independently freely choose any
one of the 40 triples $x,y,z$, and  $b$'s prior response is independent
of $A$'s choice.

To justify MIN$^\prime$ note that $a$'s response, signaled by a spot on the
screen, has already happened in an $A$-first frame, and cannot be
altered by the later free choice of $w$ by $B$; a similar remark
applies to $b$'s response. In [7], Tumulka apparently accepts this
justification for MIN$^\prime$ in rGRWf: ``$\ldots$ the first flash 
$f_A$ does not depend on the field $F_B$ in a frame in which the 
points of $B$ are later than those of $A$."

This weakening of MIN allows us to prove a weaker form of the FWT:

FWT$^\prime$:  The axioms SPIN, TWIN, and MIN$^\prime$ imply that 
there exists an inertial frame such that the response of a spin 1
particle to a triple experiment is not a function of properties 
of that part of the universe that is earlier than the response with 
respect to this frame.

This result follows without change from our present proof of the FWT
by taking $F$ to be an $A$-first frame and $G$ a $B$-first frame, and
applying MIN$^\prime$ in place of MIN to eliminate $\theta^F_a$'s 
dependence on $w$ and $\theta^G_b$'s dependence on $x,y,z$.

We can now apply FWT$^\prime$ to show that rGRWf's first flash 
function ($f_y^\Lambda$ of [4]), which determines $a$'s response, 
cannot exist, by choosing $\Lambda$ to be the frame named in FWT$^\prime$.

The Free Will Theorem thus shows that any such theory, even if it
involves a stochastic element, must walk the fine line of predicting
that for certain interactions the wave function collapses to some
eigenfunction of the Hamiltonian, without being able to specify which
eigenfunction this is. If such a theory exists, the authors have no 
idea what form it might take.}

\normalsize{
%\noindent {\bf References}

}


\begin{thebibliography}{9}

\bibitem[1]{1}  J. Conway and S. Kochen, 
\newblock The Free Will Theorem, Found.
Phys. {\bf 36} (2006), 1441-1473.

\bibitem[2]{2}  S. Kochen and E. Specker, 
\newblock The Problem of Hidden Variables in Quantum
Mechanics, J. Math. Mech. {\bf 17} (1967), 59-88.

\bibitem[3]{3}  A. Bassi and G. C. Ghirardi, 
\newblock The Conway-Kochen argument and relativistic
GRW models, Found. Phys. {\bf 37 (2)} (2007), 169-185.

\bibitem[4]{4}  D. Bohm, Phys. Rev. {\bf 85} (1952), 166 + 180 pp.

\bibitem[5]{5}  G. C. Ghirardi, A. Rimini, and T. Weber, 
\newblock Unified dynamics for microscopic and macroscopic systems,
 Phys. Rev. D34, (1986), p.470.

\bibitem[6]{6}  R. Tumulka, arXiv:0711.oo35v1 [math-ph] 31 Oct 2007.

\bibitem[7]{7}  R. Tumulka, 
\newblock Comment on ``The Free Will Theorem", Found. Phys.
{\bf 37 (2)} (2007), 186-197.

\end{thebibliography}
\end{document}